\documentclass{article}
\usepackage{graphics}
\usepackage{pstricks,pst-node,pst-tree,pst-ellipse}
\usepackage{xspace}
\usepackage{amsmath}
\usepackage{amssymb}
\newcommand{\exact}[1]{\widehat{#1}}
\newcommand{\histo}{{\mathcal P}}
\newcommand{\histobar}{{\exact{\mathcal P}}}

\newcommand{\half}{\frac{1}{2}}

\newcommand{\CC}{{\mathcal C}}
\newcommand{\Cross}{{\mathsf X}}
\newcommand{\Moebius}{{\infty}}

\newcommand{\elabel}[1]{\label{eq:#1}}
\newcommand{\eref}[1]{(\ref{eq:#1})}
\newcommand{\Eref}[1]{Eq.~(\ref{eq:#1})}
\newcommand{\flabel}[1]{\label{fig:#1}}
\newcommand{\fref}[1]{Fig.~\ref{fig:#1}}
\newcommand{\Fref}[1]{Fig.~\ref{fig:#1}}
\newcommand{\SetZet}{{\mathbb Z}}
\newcommand{\etal}{{\it et~al. }}
\newcommand{\Ssite}{{\textrm{\tiny(s)}}}
\newcommand{\Sbond}{{\textrm{\tiny(b)}}}
\newcommand{\psite}{p^\Ssite}
\newcommand{\pbond}{p^\Sbond}

\newcommand{\Amplitude}{C}

\newcommand{\Ztw}{\tilde{Z}}
\begin{document}
\title{Winding clusters in percolation on the Torus and the M\"obius
strip\\
{\small Running title: ``Winding clusters in two-dimensional percolation''}}
\renewcommand{\thefootnote}{\fnsymbol{footnote}}
\author{Gunnar Pruessner\footnotemark[2] and 
Nicholas R. Moloney\footnotemark[3]}

\date{October 15, 2003}

\maketitle

\begin{abstract}
Using a simulation technique introduced recently, we study winding
clusters in percolation on the torus and the M\"obius strip for
different aspect ratios. The asynchronous parallelization of the simulation 
makes very large system and sample sizes possible. Our high accuracy results
are fully consistent with predictions from conformal field theory. 
The numerical results for the M\"obius strip and the number distribution of
winding clusters on the torus await theoretical explanation. 
To our knowledge, this study is the first of its kind.
\end{abstract}

\footnotetext[2]{Email: gunnar.pruessner@physics.org; Address:
Department of Mathematics,
Imperial College London,
180 Queen's Gate,
London SW7~2BZ,
UK}
\footnotetext[3]{Beit Fellow for Scientific Research;
Email: n.moloney@imperial.ac.uk; Address:
Condensed Matter Theory,
Blackett Laboratory,
Imperial College London,
Prince Consort Rd,
London SW7~2BW,
UK}

\section{Introduction}
Since the seminal article of Langlands \etal \cite{LanglandsPichetPouliotStAubin:1992}, 
percolation has enjoyed a renaissance in the last decade. 
Cardy's result \cite{Cardy:1992} for the probability of a crossing cluster 
on a rectangle was the first major breakthrough of conformal field theory 
in critical percolation. Related to this work, Pinson calculated the probability 
of clusters of particular winding numbers in \cite{Pinson:1994}, 
using the partition sum for the torus formulated earlier by
di Francesco \etal \cite{diFrancescoSaleurZuber:1987}.

In 1996 Hu and Lin showed numerically \cite{HuLin:1996} that the 
probability of more than one percolating cluster is non-vanishing, 
which was proven by Aizenman \cite{Aizenman:1997} shortly afterwards. 
In fact, Cardy was able to calculate the asymptotic probability of $n$ distinct, 
simultaneously crossing clusters \cite{Cardy:1998} exactly. 
One expects similar behavior for the number of distinct, 
simultaneously winding clusters on the torus.

In this article we present numerical results for winding clusters on the
torus and on the M\"obius strip with different aspect ratios, providing
strong numerical support for the claim of conformal invariance in two
dimensional percolation. Apart from the numerical results for $r=1$ by
Langlands \etal presented in \cite{Pinson:1994}, the only other
numerical studies of this kind, to our knowledge, is one by Ziff \etal
\cite{ZiffLorenzKleban:1999}, who measure the probability of a
cross topology for very small system sizes ($16 \times 16$) and
different twists on a torus with aspect ratio $1$, and those by Newman
and Ziff \cite{NewmanZiff:2001,NewmanZiff:2000}, who measure the
probability of winding clusters on a torus with aspect ratio $1$ 
and system sizes up to $256\times 256$ and varying occupation probability. 
Also, for the M\"obius strip we are only aware of studies
on the Ising model \cite{KanedaOkabe:2001,LuWu:2001}.

\subsection{Critical percolation on the torus}
We treat site and bond percolation as the two limiting cases of the more
general site-bond percolation: Sites are occupied with probability
$\psite$ and bonds are activated with probability $\pbond$. In site
percolation all bonds are active, while in bond percolation all sites
are occupied.  Two sites are connected if the bond between them is
active and both sites are occupied.  Two sites belong to the same
cluster if they are connected by a path along occupied sites and active
bonds. On a torus and the M\"obius strip, these paths may wind around
the lattice. By our convention, $(a,b)$ counts the number of windings in
the horizontal and vertical directions, respectively.  Vertical and
horizontal directions are fixed by the definition of aspect ratio as the
vertical circumference (waist) over the horizontal circumference
(ignoring the distortion), see \fref{drawing_hori_vert}.

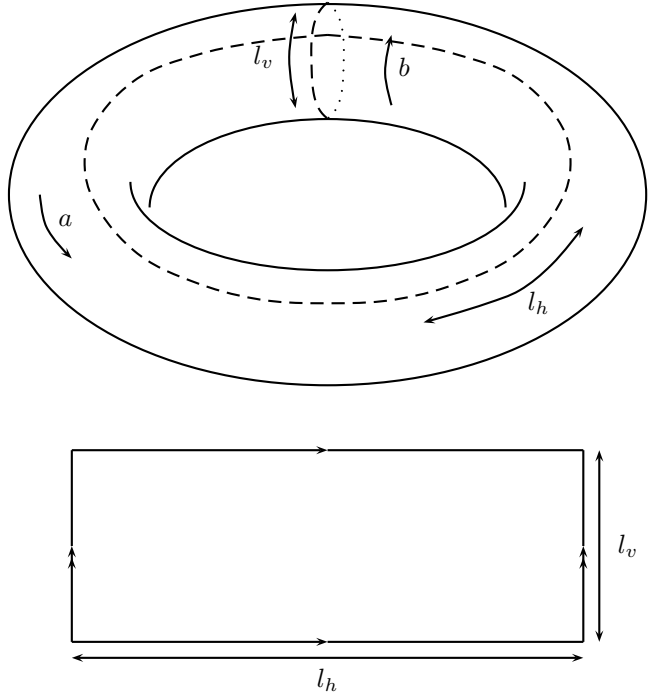
\begin{figure}[th]
\begin{center}
\begin{pspicture}(0,-4.25)(8.5,5.1)
\psset{unit=0.85cm}

\pscurve[linestyle=dashed](5,5.5)(4,5.4)(3,5.2)(2,4.8)(1.3,4)(1.2,3.5)(1.3,3)(2,2.1)(3,1.6)(4,1.35)(5,1.3)
(6,1.35)(7,1.6)(8,2.1)(8.7,3)(8.8,3.5)(8.7,4)(8,4.8)(7,5.2)(6,5.4)(5,5.5)

\pscurve[linestyle=dashed](5,6)(4.99,6)(4.75,5)(4.99,4.2)(5,4.2)
\pscurve[linestyle=dotted](5,4.2)(5.01,4.2)(5.25,5)(5.01,6)(5,6)

\pscurve{<->}(6.5,1)(7.5,1.3)(8,1.5)(8.5,1.9)(9,2.5)
\put(0,4.15){
\pscurve{<->}(4.5,1.7)(4.4,1)(4.5,0.2)
\rput(4,1){$l_v$}
}
\rput(8.3,1.3){$l_h$}

\psline{->}(1,-1)(5,-1)
\psline(5,-1)(9,-1)
\psline{->}(1,-4)(5,-4)
\psline(5,-4)(9,-4)
\psline{->>}(1,-4)(1,-2.5)
\psline(1,-2.5)(1,-1)
\psline{->>}(9,-4)(9,-2.5)
\psline(9,-2.5)(9,-1)
\psline{<->}(1,-4.25)(9,-4.25)
\psline{<->}(9.25,-4)(9.25,-1)

\rput(5,-4.6){$l_h$}
\rput(9.7,-2.5){$l_v$}

\pscurve{->}(0.5,3)(0.6,2.5)(1,2)
\pscurve{->}(6,4.4)(5.9,4.9)(6,5.5)

\rput(0.9,2.6){$a$}
\rput(6.2,5){$b$}

\psellipse(5,3)(5,3)
\psEllipticArcN(5,3.2)(3.1,1.4){0}{180}
\psEllipticArc(5,2.8)(2.8,1.4){0}{180}

\end{pspicture}
\caption{A torus with $r<1$. The winding directions of $(a,b)$ are marked as such
 and the seam is indicated as dashed line. Below, the ``unwrapped''
 lattice shows the definition of the aspect ratio, $r=l_v/l_h$.
\flabel{drawing_hori_vert}
}
\end{center}
\end{figure}

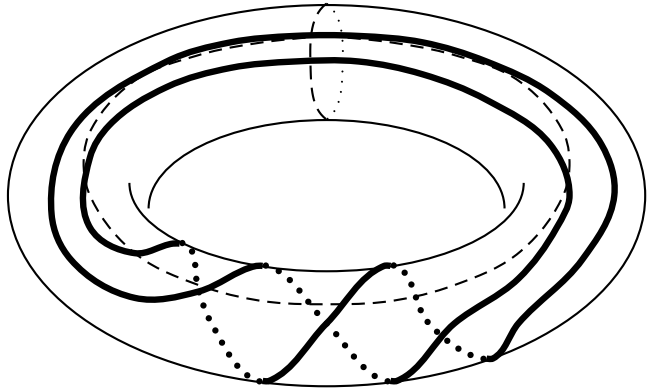
\begin{figure}[th]
\begin{center}
\begin{pspicture}(0,0)(8.5,5.1)
\psset{unit=0.85cm}
\pscurve[linewidth=0.1](4,1.9)(3.9,1.9)(3,1.5)(2,1.4)(1,2)(0.7,2.7)(1,4)(2.5,5.1)(3.5,5.4)(4.5,5.5)(5.5,5.5)(6.5,5.4)(7.5,5.1)(8.5,4.6)(9.3,3.8)
(9.5,3)(9,2)(8,1)(7.6,0.45)(7.5,0.45)

\pscurve[linewidth=0.1](6,0.1)(6.1,0.1)(7,1)(8,1.7)(8.6,2.5)(8.8,3)(8.5,3.8)(7.5,4.5)(6.5,4.9)(5.5,5.1)(4.5,5.1)(3.5,5)(2.5,4.7)(1.6,4.1)(1.3,3.6)
(1.3,2.5)(2,2.1)(2.6,2.25)(2.7,2.25)

\pscurve[linewidth=0.1](4,0.1)(4.1,0.1)(5,1)(5.9,1.9)(6,1.9)
\pscurve[linestyle=dotted,linewidth=0.1](4,1.9)(4.1,1.9)(5,1)(5.9,0.1)(6,0.1)
\pscurve[linestyle=dotted,linewidth=0.1](6,1.9)(6.1,1.9)(6.6,1)(7.4,0.45)(7.5,0.45)
\pscurve[linestyle=dotted,linewidth=0.1](4,0.1)(3.9,0.1)(3.2,1)(3,1.5)(2.8,2.25)(2.7,2.25)

\pscurve[linestyle=dashed](5,5.5)(4,5.4)(3,5.2)(2,4.8)(1.3,4)(1.2,3.5)(1.3,3)(2,2.1)(3,1.6)(4,1.35)(5,1.3)
(6,1.35)(7,1.6)(8,2.1)(8.7,3)(8.8,3.5)(8.7,4)(8,4.8)(7,5.2)(6,5.4)(5,5.5)

\pscurve[linestyle=dashed](5,6)(4.99,6)(4.75,5)(4.99,4.2)(5,4.2)
\pscurve[linestyle=dotted](5,4.2)(5.01,4.2)(5.25,5)(5.01,6)(5,6)


\psellipse(5,3)(5,3)
\psEllipticArcN(5,3.2)(3.1,1.4){0}{180}
\psEllipticArc(5,2.8)(2.8,1.4){0}{180}

\end{pspicture}
\caption{
A $(2,3)$ winding on a torus with aspect ratio $r<1$. The seam is shown
 as a dashed line.
\flabel{drawing_2_3}
}
\end{center}
\end{figure}

On the torus, it is a topological fact that if $a=0$ ($b=0$) then the only path 
which is not homotopic to a point and does not intersect itself has $b=1$ ($a=1$).
There is no topological difference between $(a,b)$ and $(-a,-b)$ --- the direction
in which these paths are taken is simply inverted. If $a\ne 0$ and $b\ne 0$
then $a$ and $b$ must be relative prime if the path does not intersect
itself. \Fref{drawing_2_3} shows an example for $(2,3)$, other examples
can be found in \cite{diFrancescoSaleurZuber:1987}. 

In the following, winding numbers $(a,b)$ will be called ``normalized'',
if $a=1$ for $b=0$, $b=1$ for $a=0$, $a$ and $b$ are relative prime if
$a\ne0$ and $b\ne0$, and
$a\ge 0$.

In addition to the above, we also mention the path with the so-called
``cross topology'' \cite{diFrancescoSaleurZuber:1987}. Such a path is 
produced by intersecting a $(1,0)$ and a $(0,1)$ path.
The seam shown in \fref{drawing_hori_vert} as a dashed line forms a path 
with a cross topology.

To transfer the notion of winding from paths to
clusters, one considers all (topologically different) paths within a
cluster. If there is a path with non-zero winding numbers that
crosses only paths which are homotopic to a point or have the same
winding numbers, the cluster is assigned the winding numbers of the
path. If there is no winding  path at all, then this cluster itself is
said to be homotopic to a point. It transpires that in any other case the
cluster contains a cross-topological path and the cluster is then said
to have a cross topology itself.

The \emph{universal} probability to obtain $n$ clusters with winding
numbers $(a,b)$ at aspect ratio $r$ is denoted in the following by
$\histobar((a,b),n,r)$. Pinson has derived the formula
\cite{Pinson:1994}
\begin{eqnarray} \elabel{pinson_a_b}
&& \histobar((a,b),\ge 1,r)= 
\sum_{l\in \SetZet} Z_{a3l,b3l}\left(\frac{2}{3}; r\right) - \\
&&\half \sum_{l\in \SetZet} Z_{a(3l+1),b(3l+1)}\left(\frac{2}{3}; r\right) - 
\half \sum_{l\in \SetZet} Z_{a(3l+2),b(3l+2)}\left(\frac{2}{3}; r\right) - \nonumber \\
&&\sum_{l\in \SetZet} Z_{a2l,b2l}\left(\frac{2}{3}; r\right) +
\sum_{l\in \SetZet} Z_{a(2l+1),b(2l+1)}\left(\frac{2}{3}; r\right) \nonumber
\end{eqnarray}
where
\begin{equation}
 Z_{m,n}(g;r) = \frac{\sqrt{g}}{\sqrt{r}\, \eta^2(e^{-2 \pi r})} \exp\left(-\pi g \big(\frac{m^2}{r}+n^2r\big)\right)\ ,
\end{equation}
and
\begin{equation}
 \histobar((a,b), \ge n, r) = \sum_{i=n}^\infty \histobar((a,b), i, r) \ .
\end{equation}
Here $\eta(q)$ is the Dedekind eta function
\begin{equation}
 \eta(q) = q^{1/24} \prod_{k=1}^\infty (1-q^k) \ .
\end{equation}

Correspondingly, the probability of a cross topology is denoted by
$\histobar(\Cross,r)$, and the probability that all clusters in a given
configuration are homotopic to a point is denoted by $\histobar(0,r)$.
The exact expressions for these probabilities are
\begin{equation} \elabel{pinson_0}
 \histobar(\Cross,r)=\histobar(0,r)=\half \left(Z_c\left(\frac{8}{3},1; r\right) - Z_c\left(\frac{8}{3},\half; r\right)\right)
\end{equation}
with
\begin{equation}
 Z_c(g,f;r) = f \sum_{m,n \in \SetZet} Z_{fm,fn}(g;r) \ .
\end{equation}

In the following, we will distinguish exact results from 
numerical results by putting a hat on all exact quantities.
Where necessary, the numerical results will also have an index 
indicating the system size and a superscript $\Ssite$ for site percolation 
and $\Sbond$ for bond percolation. For example, $\histo^\Sbond_{N=3000^2}((1,2),1,9)$ 
is the fraction of cases in which we observed a single $(1,2)$ cluster 
in bond percolation with aspect ratio $9$ and $3000^2$ sites.

The topological considerations in this paper are mainly technically
motivated and rather heuristic. For rigorous proofs, we refer to the
standard literature \cite{Mendelson:1975}.

Multiple, distinct clusters with the \emph{same} winding number can
coexist without intersecting \cite{diFrancescoSaleurZuber:1987}.  This,
however, does not apply to a cluster with a cross topology: there can
only be one such cluster on a torus. As
explained below in detail, winding clusters with incommensurable winding
numbers cannot coexist.  Thus an entire configuration of the lattice on
the torus is characterized by the winding numbers of the winding
clusters, if any, and their total number.

The M\"obius strip is a little more complicated in this respect. First of all, 
since the M\"obius strip is a non-orientable surface, a reasonable definition
of a spanning cluster connecting one side to the other is not possible, unlike, 
for example, clusters on a cylinder connecting top and bottom \cite{MoloneyPruessner:2003b,Cardy:1998}.
Winding clusters behave rather surprisingly. A single winding cluster, 
winding around only once, is possible. Meanwhile, if two winding clusters coexist, 
then at least one of them must have winding number $2$. 
In general, $n$ winding clusters require at least $n-1$ clusters with winding number $2$. 
In fact, all winding clusters are fully defined by the \emph{total} number $a$ of windings
alone: If $a$ is even, there are $a/2$ winding clusters with winding number $2$. 
If $a$ is odd, there are $(a-1)/2$ winding clusters with winding number $2$, 
and $1$ cluster with winding number $1$. \Fref{drawing_moebius} shows a M\"obius strip 
with $2$ winding paths adding up to a total winding number $3$. 
The aspect ratio of the M\"obius strip is defined as the width of the band
over the path length along the band, as shown in \fref{drawing_moebius}. 

\begin{figure}[th]
\begin{center}
\begin{pspicture}(0,-4.25)(8.5,5.3)
\psset{unit=0.85cm}
\pscurve(4.9,1.7)(3,2.4)(2,2.5)(1,3)(0.5,4)(1,4.6)(2,5)(3,5.2)(4,5.3)
(5,5.35)(6,5.3)(7,5.2)(8,5)(9,4.6)(9.5,4)(9,3)(8,2.5)(7,2.4)(3,0.8)(2,0.6)(1,1.1)(0.5,1.8)

\pscurve(9.5,1.8)(9,1.1)(8,0.6)(7,0.8)(5.15,1.6)

\pscurve(1.4,2.8)(2,3)(3,3.2)(4,3.4)(5,3.5)(6,3.4)(7,3.2)(8,3)(8.6,2.8)

\psline(0.5,4)(0.5,1.8)
\psline(9.5,4)(9.5,1.8)

\pscurve[linewidth=0.05](5,5.15)(4,5.1)(3,5)(2,4.7)(1,4.2)(0.5,3.5)(1,2.6)(2,2.2)(3,2.1)(4,1.9)(5,1.65)
(6,1.3)(7,1)(8,1)(9,1.5)(9.5,2.2)(9,2.9)(8,3.3)(7,3.5)(6,3.7)(5,3.8)
\pscurve[linestyle=dashed,linewidth=0.05](5,4.5)(4,4.45)(3,4.3)(2,4.1)(1,3.6)(0.5,2.9)(1,2)(2,1.7)(3,1.6)(4,1.6)(5,1.65)
(6,1.6)(7,1.6)(8,1.6)(9,2)(9.5,2.9)(9,3.6)(8,4.1)(7,4.3)(6,4.45)(5,4.5)
\pscurve[linewidth=0.05](5,3.8)(4,3.7)(3,3.5)(2,3.3)(1,2.9)(0.5,2.2)(1,1.5)(2,1.1)(3,1.1)(4,1.4)(5,1.65)
(6,1.8)(7,1.9)(8,2.1)(9,2.6)(9.5,3.5)(9,4.1)(8,4.7)(7,5)(6,5.1)(5,5.15)

\psline{<->}(0.25,4)(0.25,1.75)
\pscurve{<->}(2,5.2)(3,5.4)(4,5.5)(5,5.55)(6,5.5)(7,5.4)(8,5.2)

\rput(0,2.9){$w$}
\rput(5,5.8){$l$}

\psline(1,-1)(9,-1)
\psline(1,-4)(9,-4)
\psline{->}(1,-4)(1,-2.5)
\psline(1,-2.5)(1,-1)
\psline{->}(9,-1)(9,-2.5)
\psline(9,-2.5)(9,-4)
\psline{<->}(1,-4.25)(9,-4.25)
\psline{<->}(9.25,-4)(9.25,-1)

\rput(5,-4.6){$l$}
\rput(9.5,-2.5){$w$}

\end{pspicture}
\caption{
A M{\"o}bius strip with two non-intersecting winding clusters, one with winding number $2$
 (full line) and one with winding number $1$ (dashed line). Below, the ``unwrapped'' lattice
shows the definition of the aspect ratio, $r=w/l$.
\flabel{drawing_moebius}
}
\end{center}
\end{figure}
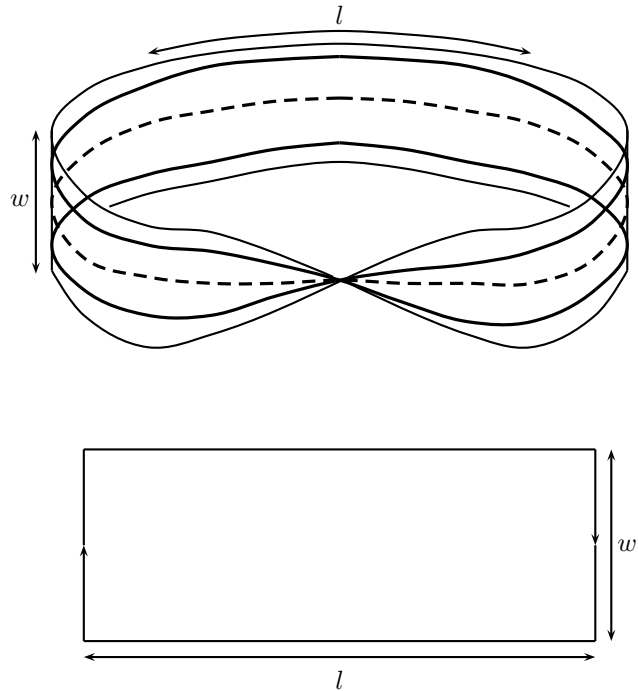

One might be tempted to extend the study of winding numbers to the Klein
bottle.  In this case, however, the wrapping around the waist is
ill-defined, just like for the spanning cluster on the M\"obius strip
discussed above. The twist in the boundary conditions breaks
translational invariance; for example, a winding path, for which the
algorithm (see below) would indicate winding numbers $(1,0)$, can be
smoothly transformed into a path with winding numbers $(1,1)$. This also
makes the Klein bottle fundamentally different from the torus with a
twist, which has a fixed ``offset'' for its winding numbers. Therefore,
we did not study the Klein bottle.

\section{Method}
The simulation is based on a method described in detail in
\cite{MoloneyPruessner:2003}, which in principle relaxes all the
standard constraints in numerical simulations of percolation, such as
CPU power, memory, and network capacity. The method is especially suited for
calculating cluster size distributions and crossing probabilities. 
Here we have used the method to calculate probabilities of winding clusters
on the torus and the M\"obius strip.

The algorithm runs asynchronously across a cluster of computers (slaves), 
which supply a central node (master) with ``patches''. 
These patches are produced by simulating percolation on squares and encoding
in the boundary information about all clusters touching the boundary. 
The data representation is due to Hoshen and Kopelman \cite{HoshenKopelman:1976} 
(HK), and the reduction of the data to the boundary is essentially a
form of ``Nakanishi label recycling'' \cite{NakanishiStanley:1980,BinderStauffer:1987}. 
The patches are ``glued together'' by the master node to form larger lattices with various
boundary conditions, such as periodic boundaries used here.
One of the key features of the algorithm is that the same patches can be
used for different topologies (such as the torus and the M\"obius strip) and aspect ratios. 
To reduce correlations between different results, patches can be randomly permuted, 
mirrored and rotated. We have applied this technique for each aspect ratio.

Technical details concerning the method can be found in \cite{MoloneyPruessner:2003}
and, especially regarding the detection of wrapping clusters, in \cite{MoloneyPruessner:2003b}.
Here we assume that a master node has received a number of square patches of size $L^2$,
and has (randomly) glued them together to form a lattice with open boundaries of size $N$ 
and aspect ratio $r$. At this stage, sites are still encoded in HK style: 
Every site on the border has a pointer attached to it, which points either to its 
representative or to itself. If it points to itself, the site is called a root site. 
A representative may or may not point to another representative, so that the pointers form a
tree-like structure. The algorithm ensures that a tree has exactly one
root and that all sites within the same cluster belong to the same
tree. Thus, all information about a cluster can be stored at the root
site and is accessible from every site belonging to the cluster.

To facilitate the calculation of winding numbers, the HK representation 
is extended so that each pointer includes a distance, indicating the multiple 
of $\pi$ from a given site to its representative. The path length from a given 
site to its root is then given by the sum over all distances along the path in 
the tree of representatives. Details about this can be found in \cite{MoloneyPruessner:2003b}.

\subsection{Identifying clusters}
When a torus is formed by the master from a rectangular lattice with
open boundaries, periodic boundaries are applied across opposite sides
of the rectangle. We call these boundaries the ``seam'' of the torus,
which is shown in \fref{drawing_hori_vert}. The seam itself is a cross
topology, so that it necessarily crosses all clusters which are not
homotopic to a point. Therefore, all winding clusters are encountered
during the gluing procedure and it is only there and then that one has
to detect winding numbers. The seam is fastened like a zip that runs
along pairs of sites. During this process clusters are detected to be
winding and/or merge. If two sites either side of the seam are seen to be
part of the same cluster prior to local gluing, because they are
connected via another path (which may or may not cross the seam closed
so far), a winding path has been detected, provided that the paths from
the two sites to the same root differ. After the new path is normalized,
two distinct cases need to be distinguished: 1) The cluster so far
contains no winding path.  In this case, the winding numbers are
assigned to the cluster.  2) The cluster is already known to be winding,
with winding numbers $(a,b)$.  If the new path $(a',b')$ obeys $(a',b')
= (a,b)$, then the new path ``runs parallel'' to the old one.  If the
old and new paths differ, the cluster \emph{must} have a cross topology:
since both paths are normalized, they cannot be multiples of each other,
because if there was a $k$ such that $(ka,kb)=(a',b')$ or
$(ka',kb')=(a,b)$ one of the two paths would not be normalized. If
$(a,b)=(1,0)$ then the new path must have $b'\ne 0$.  Therefore $(a,b)$
intersects $(a',b')$ thereby providing a ``shortcut'' and forming a path
$(0,1)$. That is, $(a',b')$ taken forwards and $(1,0)$ taken $a'$ times
backwards gives $(0,b')$, which, normalized, is just $(0,1)$ (see
below).

The general case is that both paths wind in both directions. They are
not multiples, so $a/a'\ne b/b'$. But then it is possible to find a path
of the form $(q,0)$ with $q\ne0$, namely by taking $b'$ times path
$(a,b)$ and $b$ times path $(a',b')$ backwards, which results in a path
$(b'a-ba',0)$, with $b'a-ba'\ne 0$. All paths constructed by forward and 
backward movements may have their clusters deformed in such a way that the 
resulting path is composed of only forward movements. A path $(q,0)$ with 
$|q|>1$ must intersect itself, so that this path will always contain a path $(1,0)$.

The same construction provides a path $(0,1)$. Thus once again we arrive 
at the cross topology: if a cluster contains two incommensurable paths, 
then such a cluster has a cross topology.

The recipe above provides a unique way to determine the winding
numbers of a single cluster whenever different paths are identified in it.
Similar principles apply to the entire configuration: 
The configuration itself is assigned winding numbers and a counter, 
keeping track of how often a particular type of winding cluster occurs. 
A case not mentioned above is that of two clusters, already known to be winding, 
that are merged across the seam. If the paths have the same normalized winding
numbers, the counter is reduced by one. It cannot happen that they do
not have the same winding numbers: It is topologically impossible that
two clusters are both known to be winding with different winding numbers
before being merged at a later stage. Thus, if
the first winding cluster is found, its winding number is assigned to
the entire configuration and the counter is set to one. If a further,
distinct cluster is found to be winding, it must wind with the same
winding numbers, so that the counter is increased by one. If any cluster
is found to have a cross topology, then all clusters will have a cross
topology or be homotopic to a point. The cross topology is assigned
to the entire configuration and the counter is set to one.

Similar rules can be found for the M\"obius strip. As explained above, a
single cluster with winding number $1$ can coexist with multiple
clusters of winding number $2$. Any new winding cluster found increases
the total winding number of the configuration by its winding number. If
two clusters merge, at least one of them must wind twice, reducing the
total number of windings by $2$.

\section{Results}
In this section we first list the key parameters of the simulation. Then
we discuss the assumptions and estimates of the numerical errors
associated with the study. In the following subsection we present the
results for the torus which can be compared to Pinson's formulae
\eref{pinson_a_b} and \eref{pinson_0}, asymptotes for which are
presented in Section~\ref{sec:asymptotes}. In the following
subsections, we present the results for the number distribution of
multiple winding clusters and the results for the M\"obius strip.

In each simulation, we have determined $\histo_N((a,b),n,r)$ as well as
$\histo_N(0,r)$ and $\histo_N(\Cross,r)$ for site and bond percolation
at the critical point, i.e. occupation probability $\psite=0.59274621$
\cite{NewmanZiff:2000} in site percolation and activation probability
$\pbond=1/2$ \cite{Kesten:1980} in bond percolation.

Each simulation, parametrized by the system size $N$ and the type of
percolation (site or bond), consists of at least $10^6$ realizations. A
realization requires $900$ quadratic patches of size $L^2$
($L=10,100,1000$) produced by the slaves, which are joined together by
the master, to form a lattice of size $N=900 L^2$ and aspect ratio $r$,
so that $N=30000^2, 3000^2, 300^2$. The resulting lattice is then
``glued'' together in different ways to form the different
topologies. The $14$ different aspect ratios are: $30/30$, $36/25$,
$45/20$, $50/18$, $60/15$, $75/12$, $90/10$, $100/9$, $150/6$, $180/5$,
$225/4$, $300/3$, $450/2$ and $900/1$. While a torus with aspect ratio
$r$ is topologically identical to a torus with aspect ratio $1/r$, a
M\"obius strip can be glued along two different borders to form a band
either of aspect ratio $r$ or $1/r$. Thus, $27$ different aspect ratios are
available for the M\"obius strip. 

The same $900$ patches are used for all different topologies and aspect
ratios, but they are randomly permuted, rotated and mirrored before
forming a new aspect ratio. The random number generator used is described 
in \cite{MatsumotoNishimura:1998b}, and is especially suitable for parallel 
applications.

The slaves, producing the $900$ patches, themselves apply the appropriate 
boundary conditions, so that the simulation produces of the order of $10^9$ 
samples for the torus and the M\"obius strip for $r=1$ and $L=10,100,1000$. 
We restrict the presentation of these results to special cases.

\subsection{Numerical errors}
In a numerical simulation an estimate $p$ for the probability of the
occurrence of a particular property (such as $2$ winding clusters with
winding number $(1,3)$) is measured as the average of an indicator
function $f(\CC)$ of the configuration $\CC$, which is $1$, if the
property is found in the configuration and $0$ otherwise. As all higher
moments of $f$ are $p$, the variance of $p$ is then simply estimated as
$p-p^2$, so that the variance of the estimator of the probability is
estimated as $(p-p^2)/(N-1)$, where $N$ is the number of (independent)
realizations.

It is worth mentioning a subtlety which can give rise to a systematic
overestimation of the underestimation of the probability and vice versa:
If the variance of the estimator for the probability $p$ is estimated
from the numerical data, and this data indicates a slightly smaller
probability than the exact result, the variance is likewise estimated
slightly too small, if $p<1/2$. Therefore, the relative numerical
deviation of the numerically estimated probability from the exact result
will seem slightly too large. This becomes immediately clear if one
wants to calculate the relative error for the probability for a
configuration which actually never occurred: the estimated probability
and the estimated variance both vanish. Moreover, we note that the
estimation of the variance itself acquires an increasing error as the
probability approaches $0$ or $1$.

Vice versa, if the probability $p<1/2$ is slightly overestimated from
the numerical data, so too is the variance, and therefore the
relative error is underestimated. Correspondingly for $p>1/2$, where
overestimation leads to underestimation of the variance and therefore to
overestimation of the relative error and vice versa. However, since the
aim of this work is to corroborate analytical work, all errors are based
on the numerical results only, wherever possible. Only if the numerics
give $0$ or $1$ is the error based on the analytical result. 
If the analytical result is not available because of numerical convergence 
problems, no result is shown.

As mentioned above, the same $900$ patches are used multiple times to
form rectangular lattices of different aspect ratios. Three different
sets of boundary conditions (one for the torus, two for the M\"obius
strip) are then applied to the same rectangular lattice and the
different measurements are taken. Thus, there are correlations which,
however, we do \emph{not} explicitly account for, primarily because we 
expect them to be very small. In fact, the correction to the error indicated 
depends on how many measurements are considered simultaneously:
Assuming maximum correlations in $n$ measurements, one could divide the sample 
size by $n$, as if each quantity were based on its own subset of the sample. 
Accepting the averages calculated from the complete sample as good estimators 
of the averages calculated from each subsample, this procedure leads to a 
factor of $\sqrt{n}$ in front of each standard deviation.

In the following we make use of the symmetry of the torus,
\begin{eqnarray}
 \histobar((a,b),  n, r) & = & \histobar((b,a),  n, r^{-1}) \nonumber \\ 
 \histobar(0,      n, r) & = & \histobar(0,      n, r^{-1}) \nonumber \\ 
 \histobar(\Cross, n, r) & = & \histobar(\Cross, n, r^{-1}) \nonumber \ ,
\end{eqnarray}
i.e. the results of $r$ and $1/r$ are not independent.

\subsection{Cross topology}
One surprising result in Pinson's paper \cite{Pinson:1994} is that the probability of
a cluster with cross topology (see the seam in \Fref{drawing_hori_vert}), $\histobar(\Cross, r)$, 
is identical to the probability of no winding cluster at all, $\histobar(0, r)$, 
i.e. the probability that \emph{all} clusters are homotopic to a point. 
This is in perfect agreement with our numerical results: 
For $N=30000^2$ 
\Fref{cross_point_1000}
shows the deviation of $\histo_N(\Cross, r)$ and $\histo_N(0, r)$ from
the exact result \eref{pinson_0} for site and bond percolation\footnote{There seems to be a subtle underestimation of
$\histobar(0,r)$ and an overestimation of $\histobar(\Cross,r)$, best
seen in the slight segregation of triangles and circles. 
We are not sure yet what this effect could indicate. It is not present for
$N=3000^2$ and $N=300^2$.}.

\begin{figure}[th]
\begin{center}
\scalebox{0.65}{ \includegraphics{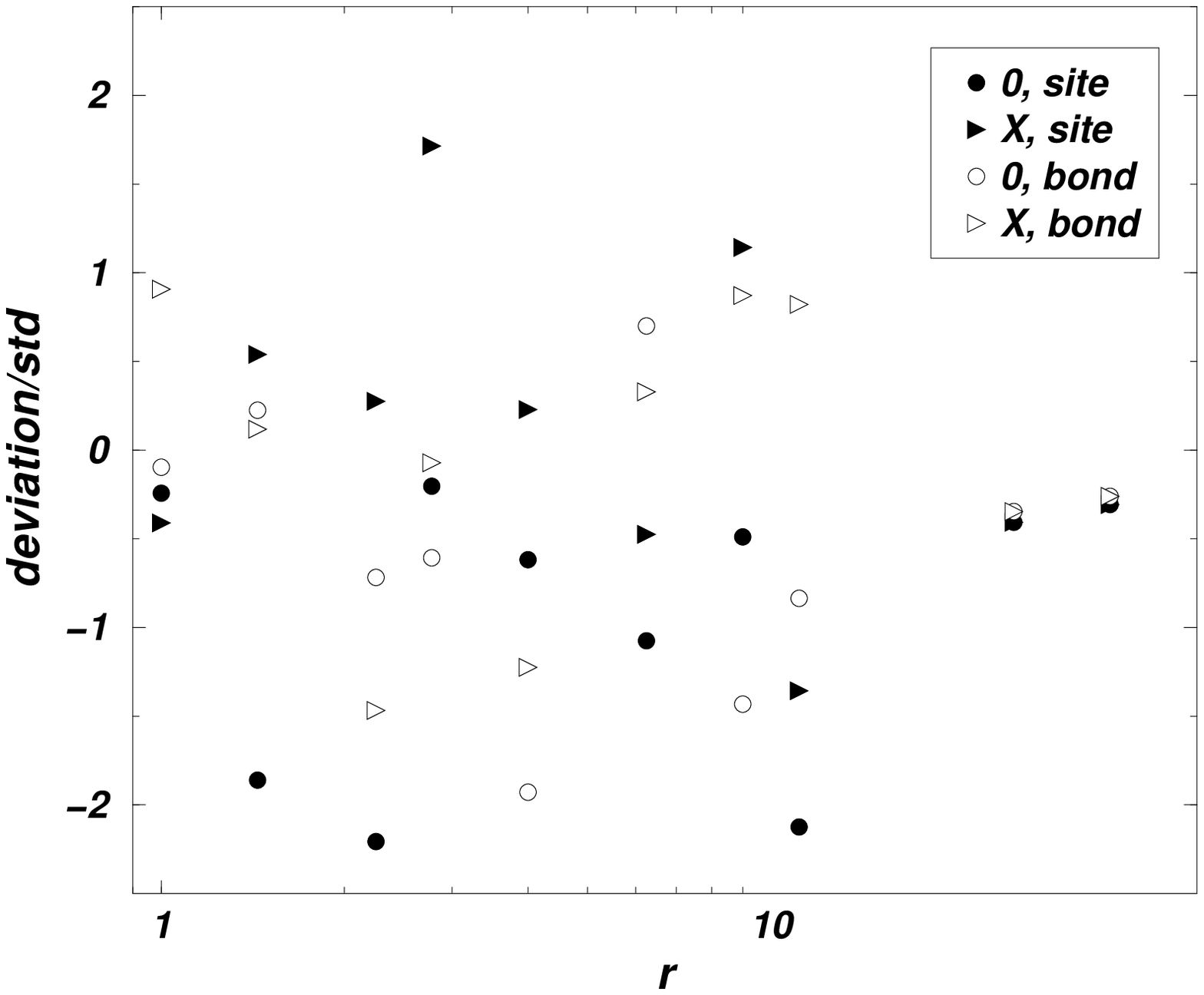}} \caption{The
 deviation of $\histo_{N=30000^2}(\Cross, r)$ and $\histo_{N=30000^2}(0, r)$ from
 $\histobar(\Cross, r)$ (see \eref{pinson_0}) in units of standard
 deviations for site (filled symbols) and bond (opaque symbols) percolation.
\flabel{cross_point_1000}
}
\end{center}
\end{figure}

It is obvious that there can only be one cluster with a
cross topology, and that any other cluster must be homotopic to a
point, i.e. if a configuration contains a cluster with a cross topology,
there is no other non-trivial cluster.

\subsection{Winding clusters} \label{sec:multi_winding}
Next we investigate the probability of at least one cluster with
winding numbers $(a,b)$, the exact expression for which was conjectured
by Pinson to be \eref{pinson_a_b}. \Fref{winding_deviation_pos} shows
the relative deviation of the numerical results from the exact value for
clusters with winding numbers $(1,0)$, $(1,1)$ and $(1,2)$ for
$N=30000^2$ and bond percolation. The reason why so many points seem to 
indicate no deviation at all is that the probability of certain types of 
winding clusters is extremely small. As a result, some rare types were not 
observed in the simulation and the resulting deviation from the exact result 
is approximately $\sqrt{p}/\sqrt{N-1}$, i.e. extremely small. 
Moreover, it should be noted that the numerical error of the evaluation of
\eref{pinson_a_b} increases as the probability approaches $0$ or $1$.

\begin{figure}[th]
\begin{center}
\scalebox{0.65}{ \includegraphics{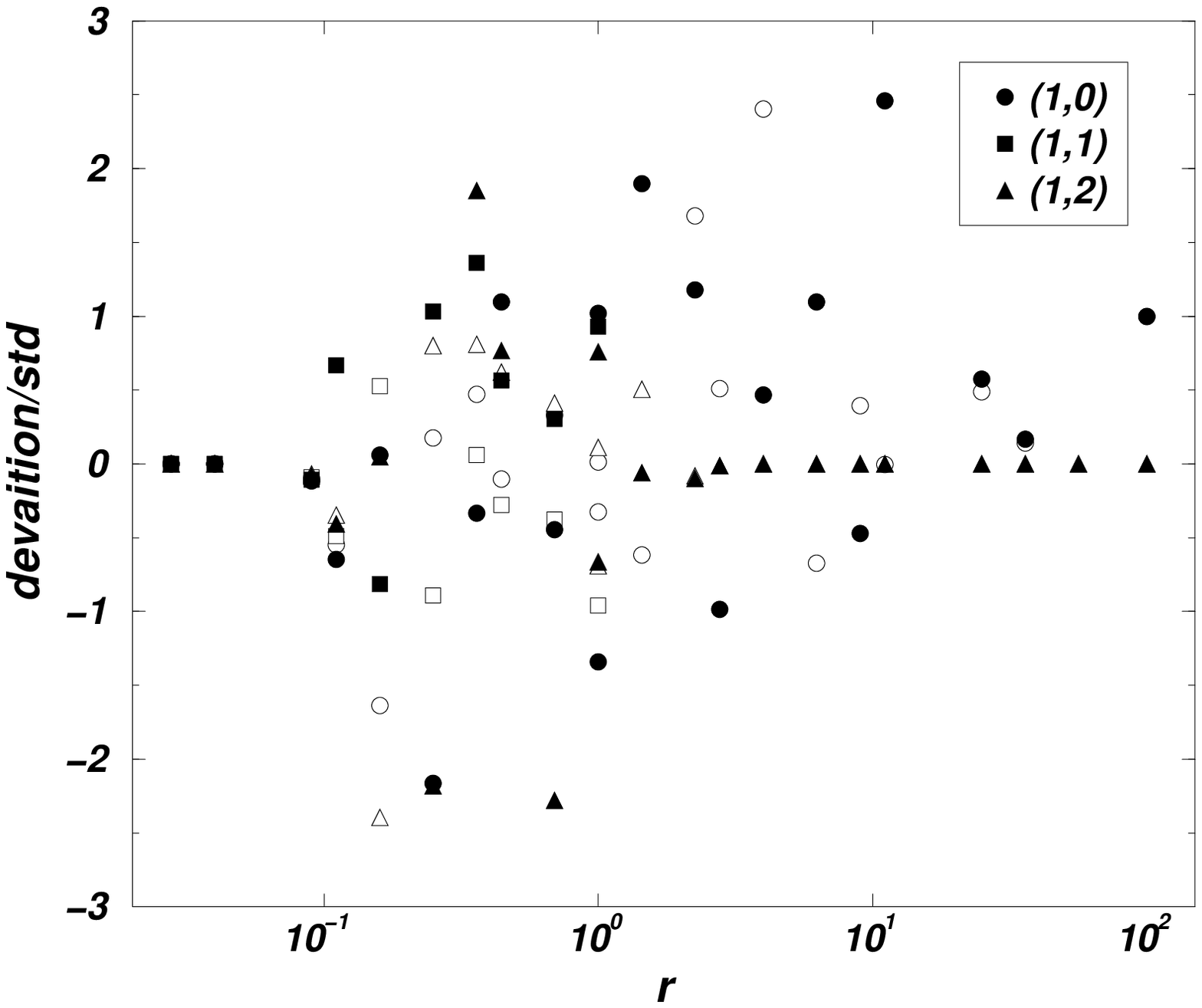}}
 \caption{The deviation of $\histo_{N=30000^2}((a,b), \ge 1, r)$ from
 $\histobar((a,b), \ge 1, r)$ (see \eref{pinson_a_b}) in units of
 standard deviations for $(a,b)=(1,0), (1,1), (1,2)$ and site (filled
 symbols) and bond (opaque symbols) percolation.
 \flabel{winding_deviation_pos} }
\end{center}
\end{figure}

\Fref{winding_1_2} shows a typical set of data for winding numbers
$(1,2)$. It is clear that the probability for a $(1,2)$ winding cannot
be symmetric around $r=1$. Above $r=2.0$ it vanishes, because it becomes
less and less likely for a cluster to wind around twice as the aspect
ratio increases. Correspondingly, as $r$ decreases it becomes less and 
less likely for a cluster to wind around once in the orthogonal direction.

All our numerical findings are fully consistent with \eref{pinson_a_b}
for winding numbers 
$(1,0)$, 
$(1,\pm 1)$,
$(1,\pm 2)$ 
and
$(1,\pm 3)$, 
in site and bond percolation.
Other winding numbers occur too rarely to make any firm statements.

\begin{figure}[th]
\begin{center}
\scalebox{0.65}{ \includegraphics{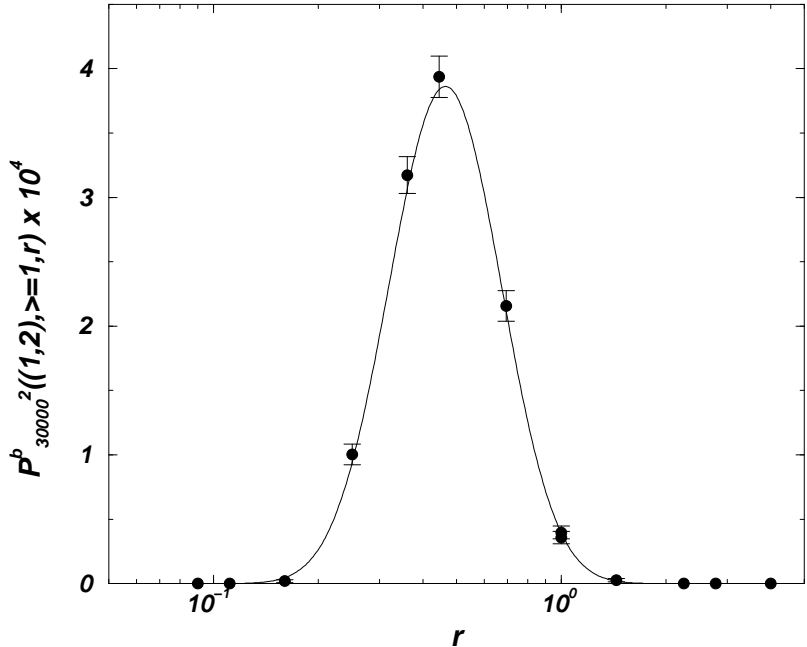}} \caption{The
 rescaled probability $\histo_{N=30000^2}^\Sbond((1,2), \ge 1, r)$ for
 different aspect ratios. The rescaling factor is $10^4$. The straight
 line shows the analytical result \eref{pinson_a_b}.
 \flabel{winding_1_2} }
\end{center}
\end{figure}

Percolation on a torus is mirror symmetrical, i.e. 
\begin{equation}
 \histobar((a,b), n,r)=\histobar((a,-b), n,r)=\histobar((-a,b), n,r)=\histobar((-a,-b), n,r) \ .
\end{equation} 
While a cluster of type $(a,b)$ is simply identical to a cluster of
type $(-a,-b)$, the windings $(a,b)$ and $(a,-b)$ are in fact
distinguishable. Since these two quantities have the same probability, 
their comparison affords a consistency check upon the numerics, 
independent of any theoretical result or finite size corrections.
We have passed this test successfully.

\subsection{Asymptotes} \label{sec:asymptotes}
To discuss numerical results for the asymptotic behavior 
of multiple winding clusters, we must first extract 
an appropriate functional form to fit against from \eref{pinson_a_b}. 
A calculation closely related 
to the following was presented by Ziff for the problem of crossing in 
two-dimensional percolation \cite{Ziff:1995,Ziff:1995_addendum} and 
subsequently used in \cite{MoloneyPruessner:2003b}.

It is convenient to slightly rewrite \Eref{pinson_a_b}. 
Noting that all $Z_{m,n}$ enter \eref{pinson_a_b} in the form $Z_{ak,bk}$, 
it is reasonable to define 
\begin{equation} \elabel{defzl}
 \Ztw((a,b);g,r)=\sqrt{\frac{g}{r}}\  \sum_{l\in\SetZet} \exp\left(-l^2 \pi g \big(\frac{a^2}{r} + b^2 r\big)\right)\ ,
\end{equation}
which is just $\sqrt{g/r}\, \vartheta_3(0,\tau)$, where $\tau=i g (a^2/r +b^2 r)$ 
and $\vartheta_3$ is Jacobi's theta function \cite{MagnusOberhettingerSoni:1966}.  
By using transformations of the form $\sum_l f_{2l+1}=\sum_l f_l - \sum_l f_{2l}$ one has
\begin{eqnarray} \elabel{my_a_b}
&& \histobar((a,b),\ge 1,r) = \frac{1}{\eta\big(e^{-2 \pi r}\big)^2} \\
&& \times
\left(\half \Ztw((a,b);6,r) - \Ztw((a,b);8/3,r) + \half \Ztw\big((a,b);2/3,r)  \right)  \nonumber
\ ,
\end{eqnarray}
given appropriate convergence of the sums in \eref{defzl} and \eref{pinson_a_b}.
For $(a,b)=(1,0)$ \Eref{defzl} transforms into a Riemann sum 
with mesh size $1/\sqrt{r}$, and for large $r$ it converges to a Gaussian integral,
i.e. $\lim_{r\to\infty} \Ztw((a,b);g,r) = 1$. That, however, leads to
$\lim_{r\to\infty} \exp(\pi r/6) \times 0$ in \eref{my_a_b}. 
This can easily be fixed using the Poisson summation formula (or the properties
of $\vartheta_3$ \cite{MagnusOberhettingerSoni:1966}), which gives
\begin{equation}
 \sum_{l\in\SetZet} e^{-l^2 u^2} u = \sqrt{\pi} \sum_{l\in\SetZet} e^{-\pi^2 l^2 / u^2}
\end{equation}
and therefore for general $(a,b)$
\begin{equation} \elabel{zl_inverse}
 \Ztw((a,b); g,r) = \frac{1}{\sqrt{a^2+b^2 r^2}} \sum_{l\in\SetZet} \exp\left(-\frac{\pi}{g(a^2 + b^2 r^2)}l^2 r\right) \ .
\end{equation}

From the definition of the Dedekind eta function
\begin{equation}
 \eta(\exp(-2 \pi r)) = e^{-\frac{\pi r}{12}} \prod_{k=1}^\infty \left(1-e^{-2k\pi r}\right)
\end{equation}
one finds the following expansion for large $r$:
\begin{equation} \elabel{dedekind_expansion}
 \eta^{-2}(\exp(-2 \pi r))=e^{\pi r/6} \left(1 + 2 e^{-2 \pi r} + 5 e^{-4 \pi r} + 10 e^{-6 \pi r} + \dots\right) \ .
\end{equation}
Using \eref{zl_inverse} for the large $r$ expansion of \eref{my_a_b} at
$(a,b)=(1,0)$
\begin{equation} \elabel{z10_expansion}
 \Ztw((1,0);g,r) = 1+2 e^{- \frac{\pi r}{g}} +2 e^{- 4 \frac{\pi r}{g}} +2 e^{- 9 \frac{\pi r}{g}} +2 e^{- 16 \frac{\pi r}{g}} + \dots \ ,
\end{equation}
the task boils down to ordering terms:
\begin{equation}
 \histobar((1,0),\ge 1,r)=1-2e^{-\frac{5}{24}\pi r}+e^{-\half \pi r} + 2 e^{-2 \pi r} - 4 e^{-\frac{53}{24}\pi r} + 3 e^{-\frac{5}{2}\pi r}\dots
\end{equation}
This approximation has a relative deviation from the exact result of
less than $5 \times 10^{-4}$ at $r=1$ and less than $10^{-8}$ at $r=2$.

Using $\histobar((1,0),\ge 1,r)=\histobar((0,1),\ge 1,1/r)$, the
corresponding expansion for small $r$ is now based directly on
\eref{defzl}. Thus again for large $r$
\begin{equation} \elabel{histobar_01}
 \histobar((0,1),\ge 1,r)=\sqrt{\frac{2}{3r}}\left(e^{-\half \pi r} - e^{-\frac{5}{2} \pi r} - e^{-\frac{9}{2} \pi r} + 4 e^{-\frac{35}{6} \pi r} \dots \right) \ .
\end{equation}
At $r=1$ the relative error of this approximation is better than
$3 \times 10^{-8}$, which improves to about $10^{-15}$ at $r=2$.

Similarly one finds for $\histobar(\Cross,r)$
\begin{equation}
  Z_c(f,g;r) = \frac{1}{\eta\big(e^{-2 \pi r}\big)^2} \Ztw\big((1,0);f^2g,r\big)\ \Ztw\big((1,0);1/(f^2g),r\big) 
\end{equation}
which yields together with \eref{dedekind_expansion} and \eref{z10_expansion}
\begin{equation}
 \histobar(\Cross,r)=e^{-\frac{5}{24} \pi r} - e^{-\frac{1}{2} \pi r} - 2 e^{-2 \pi r} + 2 e^{-\frac{53}{24} \pi r} - 2 e^{-\frac{5}{2} \pi r} \dots
\end{equation}
again for large $r$. The relative deviation is less than $9 \times 10^{-4}$
at $r=1$ and about $10^{-7}$ at $r=2$. 

\subsection{Multiple winding clusters}
Unfortunately, it is not straightforward to extend Cardy's
qualitative arguments in the introduction of \cite{Cardy:1998} for the
existence of multiple spanning clusters.

Multiple winding clusters with winding numbers other than $(1,0)$ are very rare. 
In fact, we found
\begin{subequations}
\begin{eqnarray}
\hspace{-1pc}
\histo^\Ssite_{N=1000^2}((1,1),2,1)+\histo^\Ssite_{N=1000^2}((1,-1),2,1)\!\!\! &=&\!\!\! 1.40(8) \times 10^{-7} \\
\hspace{-1pc}
\histo^\Sbond_{N=1000^2}((1,1),2,1)+\histo^\Sbond_{N=1000^2}((1,-1),2,1)\!\!\! &=&\!\!\! 1.50(10) \times 10^{-7}
\end{eqnarray}
\end{subequations}
based on the data produced at the slave nodes. Probabilities for higher
multiples and other winding numbers are less than about 1 in $2 \times 10^9$.

\begin{figure}[th]
\begin{center}
\scalebox{0.65}{ \includegraphics{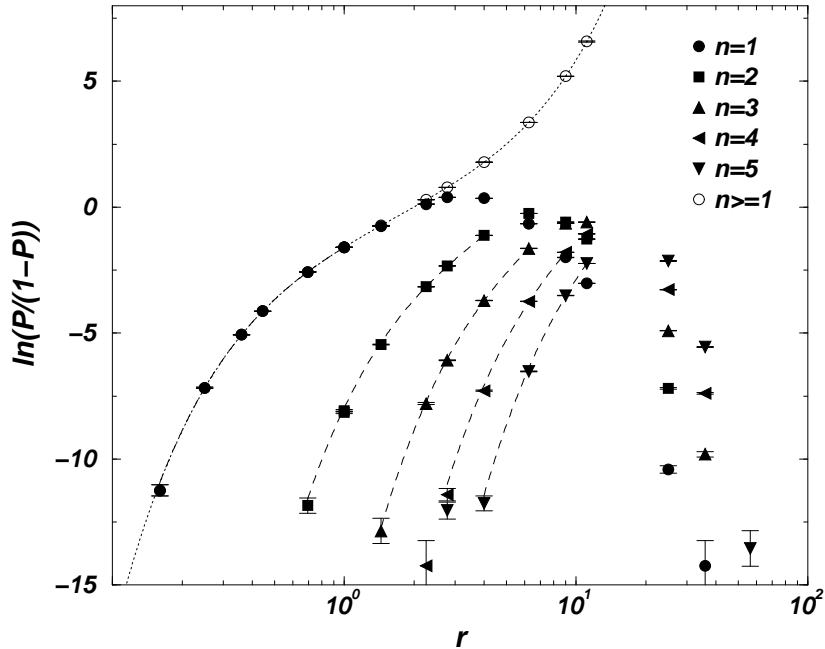}} \caption{The
 rescaled probability $\histo^\Sbond_{N=30000^2}((1,0), n, r)$ in the
 form $\ln(\histo/(1-\histo))$ for
 different $n$. The dotted line shows the analytical result
 \eref{pinson_a_b}, while the dashed lines are fitted according to
 \eref{expected_multi} using the parameters shown in
 Tab.~\ref{tab:multi_winding_fits}. 
 \flabel{multiple_windings} }
\end{center}
\end{figure}

\begin{table}
\begin{tabular}{c|l|ll|ll}
$n$          & $r$ range            & $\Amplitude^\Ssite((1,0), n)$ & $\alpha((1,0),n)$ & $\Amplitude^\Sbond((1,0),n)$ & $\alpha((1,0),n)$\\
\hline
$1$          & $12/75 \dots 30/30$  & 0.811(3)                     & 1.568(3)          & 0.810(3)                     & 1.566(3)         \\
$2$          & $25/36 \dots 60/15$  & 0.856(3)                     & 7.771(12)         & 0.868(4)                     & 7.813(14)        \\
$3$          & $36/25 \dots 75/12$  & 1.364(10)                    & 19.02(4)          & 1.368(11)                    & 19.02(5)         \\
$4$          & $50/18 \dots 90/10$  & 2.05(2)                      & 33.91(8)          & 2.10(2)                      & 34.05(10)        \\
$5$          & $60/15 \dots 100/9$  & 3.76(6)                      & 53.94(16)         & 3.71(7)                      & 53.82(19)        
\end{tabular}
\caption{\label{tab:multi_winding_fits}
Multiple winding clusters $(1,0)$. The numerical results
 $\histo_{N=30000^2}((1,0),n,r)$ are fitted within the range indicated
 against \eref{expected_multi}.
}
\end{table}

From what has been said in Section~\ref{sec:asymptotes} (see also
\Eref{convergence}) one might suspect that the probability of $n$
distinct, simultaneously winding clusters with winding number $(1,0)$
behaves in the limit of small $r$ like
\begin{equation} \elabel{expected_multi}
 \histobar((1,0),n,r) \approx \Amplitude((1,0),n) e^{-\alpha((1,0),n)/r} \sqrt{r} \ .
\end{equation}
The variable $\Amplitude((1,0),n)$ denotes the amplitude for this type of
winding and $\alpha((1,0),n)$ is expected to be a second order
polynomial in $n$ \cite{Cardy:1998}. Clearly, the additional factor $\sqrt{r}$ is ``only'' a
logarithmic correction to the dominating exponential, but is in fact 
clearly visible in the numerics. 

Nothing can be derived from Section~\ref{sec:asymptotes} concerning the large $r$ limit,
since the average number of winding clusters and the variance thereof become very large.

\Fref{multiple_windings} shows the estimated probabilities in the form
$\ln(\histo/(1-\histo))$ \cite{Cardy:1992,MoloneyPruessner:2003b}. 
These probabilities have been fitted against \eref{expected_multi} in the
small $r$ limit, the results of which are shown in Tab.~\ref{tab:multi_winding_fits}. 
Even though the choice of \eref{expected_multi} is somewhat arbitrary for $n>1$,
very good fits are obtained. These are shown by the dashed lines in \fref{multiple_windings},
which represent the fitting results of Tab.~\ref{tab:multi_winding_fits} fed back into 
\eref{expected_multi}. The main source of the numerical error is the fitting range, 
which is listed in the table. The range is bounded below by the smallest value of $r$ 
that is supported by available data within reasonable error, and bounded above by the 
approximate value for $r$ after which the asymptotic behavior terminates.
The ambiguity of the fitting range is not reflected in the error bars, which indicate 
only the statistical error. Therefore the exact result fitted to the function 
\eref{expected_multi} within the given interval should produce the values listed above.

The dotted line in \fref{multiple_windings} shows $\histobar((1,0),\ge1,r)$, 
which asymptotically (small $r$) contains only a single winding cluster, i.e. 
\begin{equation} \elabel{convergence}
 \lim_{r\to 0}\ \frac{\histobar((1,0),\ge1,r)-\histobar((1,0),1,r)}{\histobar((1,0),\ge1,r)+\histobar((1,0),1,r)} = 0
\end{equation}
so that the amplitude $\Amplitude((1,0),1)$ and the exponent $\alpha((1,0),1)$ are known
exactly from \eref{histobar_01}:
\begin{subequations}
\begin{eqnarray}
 \Amplitude((1,0),1)&=&\sqrt{2/3}=0.8164909\dots \\ 
  \alpha((1,0),1)&=&\pi/2 = 1.5707963\dots
\end{eqnarray}
\end{subequations}
which is in very good agreement with the results in Tab.~\ref{tab:multi_winding_fits}, 
indicating that the fitting range chosen there is reasonable.

Just as for wrapping clusters on the cylinder, one might be tempted to find a
systematic dependence of $\Amplitude((1,0),n)$ and $\alpha((1,0),n)$ on $n$, 
such as an exponential and a second order polynomial, respectively. 
However, we were unable to identify these functions. 
Moreover, as the functional form \eref{expected_multi} already differs from 
the corresponding function for wrapping clusters on the cylinder, 
it is not surprising that no similarities were found between their exponents 
and amplitudes.

\subsection{Finite size effects}
The system sizes used are so large ($9 \times 10^8$ sites) that one might
be inclined to completely dismiss finite size corrections. Their study 
is nevertheless interesting for two main reasons. First, it is unknown
\emph{a priori} whether a system is actually large enough for finite 
size corrections to be neglected. Second, if they can be neglected, it is 
then interesting to investigate how strong the corrections are for smaller 
system sizes.

The strength of finite size corrections has already been discussed in
\cite{MoloneyPruessner:2003} for a simulation with the same parameters,
for crossing, spanning and wrapping on the square lattice and the cylinder. 
There are, apart from Pinson's exact results, no estimates for
probabilities specific to winding clusters on the torus that we know of.
\Fref{winding_deviation_pos} cannot be used as an indicator of high
accuracy, because the aim of this paper is exactly to substantiate
Pinson's analytical work and conformal invariance in percolation as such. 

\begin{figure}[th]
\begin{center}
\scalebox{0.65}{ \includegraphics{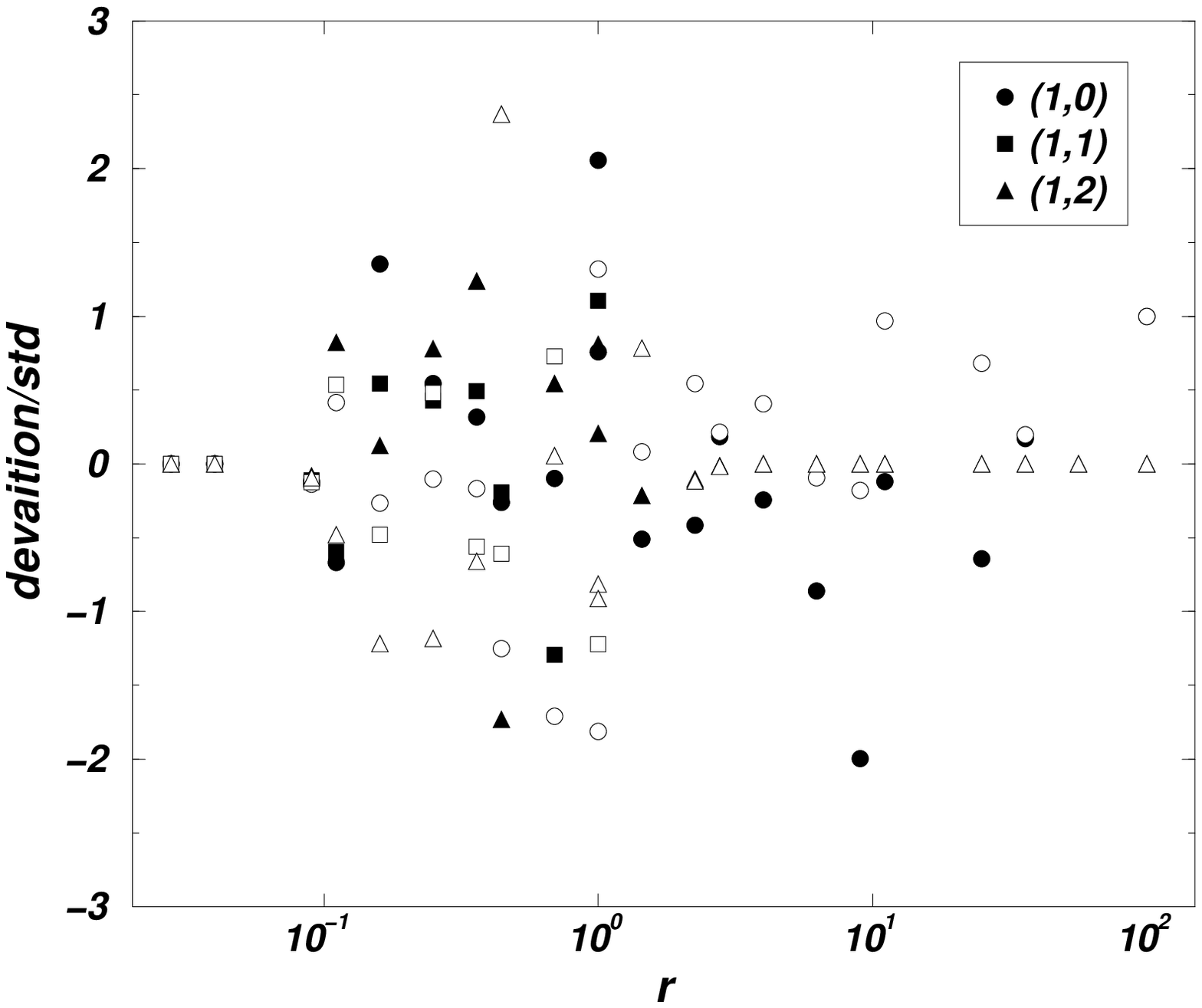}}
 \caption{The deviation of $\histo_{N=300^2}((a,b), \ge 1, r)$ from
 $\histobar((a,b), \ge 1, r)$ (see \eref{pinson_a_b}) in units of
 standard deviations for $(a,b)=(1,0), (1,1), (1,2)$ and site (filled
 symbols) and bond (opaque symbols) percolation. There is almost no
 difference to the corresponding data shown for $N=30000^2$ in
 \fref{winding_deviation_pos}. 
 \flabel{winding_deviation_pos_10} }
\end{center}
\end{figure}

However, it is very instructive to compare the numerical findings for different
system sizes. Most remarkably, for the cross topology, the configuration
homotopic to a point, as well as winding numbers $(1,0)$, $(1,1)$ and
$(1,2)$, the distributions of the deviations from the exact results look
essentially like \fref{cross_point_1000} and \fref{winding_deviation_pos} 
even for $N=300^2$. The data corresponding to the latter are shown in 
\fref{winding_deviation_pos_10}.

We note that the deviations of the site percolation results from the
exact values are not stronger than those of the bond percolation
results. Thus we have also tested the estimate for $\psite=0.59274621$ 
\cite{NewmanZiff:2000}, which was derived on much smaller system sizes.

\subsection{M\"obius strip}
The probability to obtain $n$ windings on a M\"obius strip is denoted in
the following by $\histo(\Moebius,n,r)$, where $n$ encodes the number of
winding clusters and their winding numbers as discussed above. 
Again, two different aspects can be investigated: the probability
$\histo(\Moebius,\ge 1,r)$ and the probability $\histo(\Moebius,a,r)$ for
each individual $a=1,2,\dots$. 

\begin{figure}[th]
\begin{center}
\scalebox{0.65}{ \includegraphics{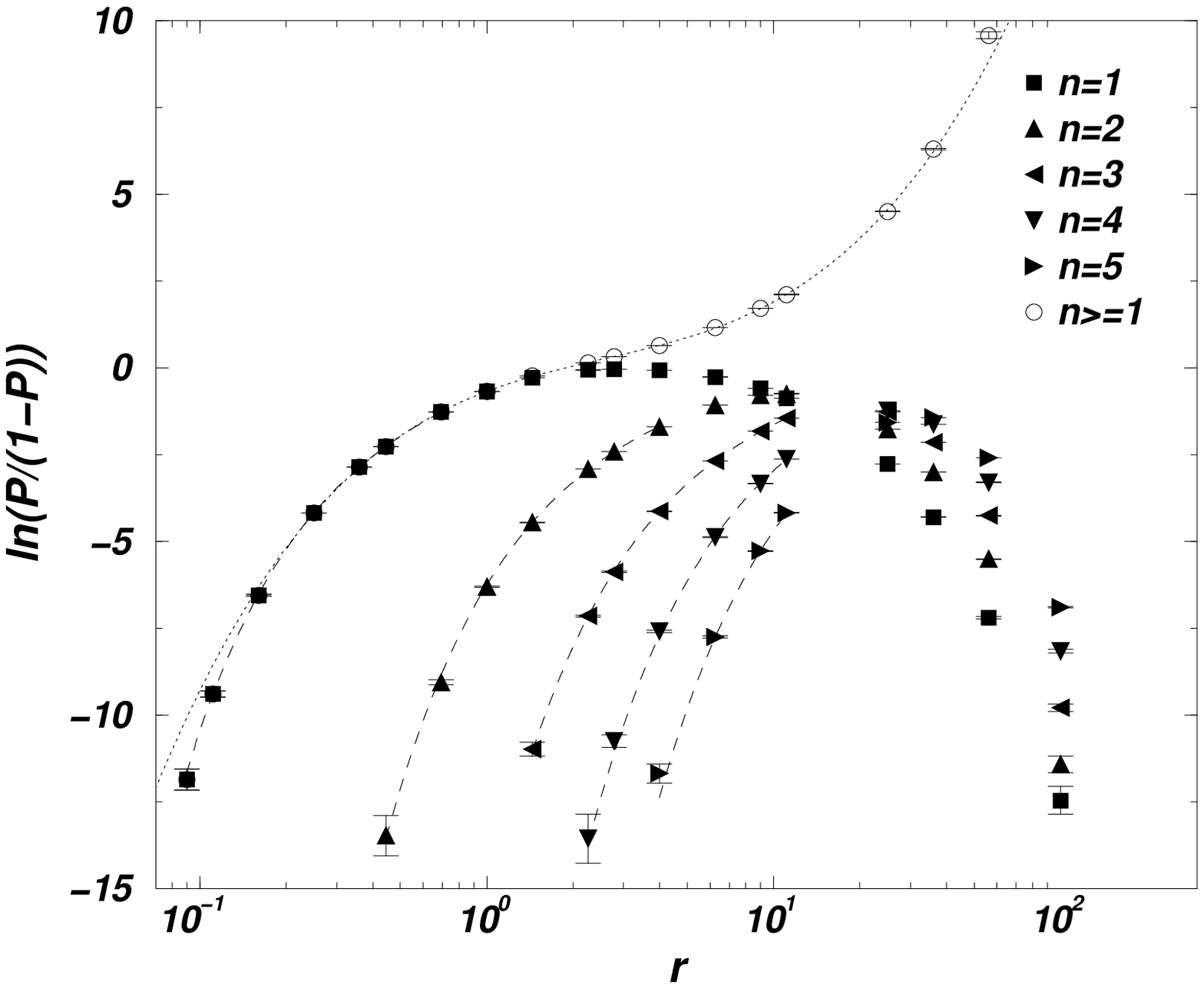}} \caption{The
 rescaled probability $\histo^\Sbond_{N=30000^2}(\Moebius, n, r)$ in the
 form $\ln(\histo/(1-\histo))$ for
 different $n$. The dashed lines show the fitting results according to
 \eref{expected_moebius} using the parameters shown in
 Tab.~\ref{tab:moebius_results}. The dotted line approximates the
 rescaled probability for $r$ around $1$ according to \eref{fit_moeb_poly_bond}.
 \flabel{moebius} }
\end{center}
\end{figure}

\Fref{moebius} shows the numerical results for bond percolation and
$N=30000^2$, again in the reduced form $\ln(\mathcal{P}/(1-\mathcal{P}))$. The behavior is
qualitatively similar to the one shown for multiple wrapping clusters on
the torus, \fref{multiple_windings}. Using the same ansatz as for
wrapping clusters on the cylinder \cite{MoloneyPruessner:2003b},
\begin{equation} \elabel{expected_moebius}
 \histobar(\Moebius,n,r) \approx \Amplitude(\Moebius,n) e^{-\alpha(\Moebius,n)/r} \ ,
\end{equation} 
the results shown in Tab.~\ref{tab:moebius_results} have been derived. 
The same precautions as in Subsec.~\ref{sec:multi_winding} apply
to the ranges shown in the table. While $\alpha(\Moebius,1)$ can be
conjectured to be $\pi/3$ with some confidence, the functional
dependence of the exponents and amplitudes on $n$ could not be found. 
Nevertheless, according to \fref{moebius} the form
\eref{expected_moebius} seems to work quite well.

\begin{table}
\begin{tabular}{c|l|ll|ll}
$n$          & $r$ range            & $\Amplitude^\Ssite(\Moebius, n)$ & $\alpha(\Moebius,n)$ & $\Amplitude^\Sbond(\Moebius,n)$ & $\alpha(\Moebius,n)$\\
\hline
$1$          & $9/100 \dots 25/36$  & 0.982(3)                     & 1.0401(14)          & 0.985(3)                     & 1.0433(17)         \\
$2$          & $25/36 \dots 60/15$  & 0.671(2)                     & 5.832(9)            & 0.676(3)                     & 5.849(11)        \\
$3$          & $36/25 \dots 100/9$  & 0.780(2)                     & 15.62(2)            & 0.774(3)                     & 15.54(3)         \\
$4$          & $45/20 \dots 100/9$  & 1.081(11)                    & 30.93(10)           & 1.099(13)                    & 31.08(12)        \\
$5$          & $60/15 \dots 100/9$  & 1.44(6)                      & 50.6(4)             & 1.51(6)                      & 51.1(4)        
\end{tabular}
\caption{\label{tab:moebius_results}
Winding clusters on the M\"obius strip; winding $n$ means $n/2$ winding
 clusters with winding number $2$, if $n$ is even, and $(n-1)/2$ clusters
 with winding number $2$ plus a single cluster with winding number $1$,
 if $n$ is odd. The numerical results
 $\histo_{N=30000^2}(\Moebius,n,r)$ are fitted within the range indicated
 against \eref{expected_moebius}.
}
\end{table}

In the spirit of \cite{LanglandsPichetPouliotStAubin:1992} we have tried
to fit $\histo_{N=30000^2}(\Moebius,\ge1,r)$ against a third order polynomial in $\ln(r)$. 
The result
\begin{subequations}
\elabel{fit_moeb_poly}
\begin{eqnarray}
&& \ln\left(\frac{\histo^\Ssite_{N=30000^2}(\Moebius,\ge1,r)}{1-\histo^\Ssite_{N=30000^2}(\Moebius,\ge1,r)} \right)  \approx  \\
&& -0.6798(7) + 1.3525(9)  \ln(r) - 0.5648(9) \ln(r)^2 + 0.2021(3) \ln(r)^3 \nonumber \\
&& \nonumber \\
&& \ln\left(\frac{\histo^\Sbond_{N=30000^2}(\Moebius,\ge1,r)}{1-\histo^\Sbond_{N=30000^2}(\Moebius,\ge1,r)} \right) \approx  \elabel{fit_moeb_poly_bond}\\
&& -0.6793(8) + 1.3545(10) \ln(r) - 0.5689(11)\ln(r)^2 + 0.2033(4) \ln(r)^3 \nonumber 
\end{eqnarray}
\end{subequations}
is shown for bond percolation in \fref{moebius} as well. Similar to the
results above, the main source of error is not statistical, but systematic, 
namely in the choice of the specific function. Nevertheless, the numerical result 
\eref{fit_moeb_poly} will possibly serve as a reference for analytical findings.

\section{Conclusion}
Based on a large scale numerical simulation, this paper provides one of
the first numerical confirmations of Pinson's analytical results for
winding clusters on the torus, which are based on conformal field
theory. It therefore also supports conformal invariance at the
critical point. 

By rewriting Pinson's results, it was possible to derive some asymptotes
that have hitherto only been derived for the flat topology
\cite{Ziff:1995,Ziff:1995_addendum}. These asymptotes have been used in
the investigation of the probability of multiple, simultaneously
wrapping clusters.

A similar numerical analysis has been carried out for the M\"obius strip,
which still awaits analytical treatment.

\section*{Acknowledgments}
The authors wish to thank Andy Thomas for his fantastic technical
support. Without his help and dedication, this project would not have
been possible.  The authors also thank Dan Moore, Brendan Maguire and
Phil Mayers for their continuous support. NRM is very grateful to the
Beit Fellowship, and to the Zamkow family. GP gratefully acknowledges
the support of the EPSRC.  
\bibliography{articles,books}
\end{document}